\documentclass[
  cha,
  aip,
  reprint,
  showpacs
]{revtex4-1}

\usepackage[T1]{fontenc}
\usepackage[utf8]{inputenc}
\usepackage{amsmath}
\usepackage{amssymb}
\usepackage{graphicx}
\usepackage{grffile}
\usepackage{refstyle}
\usepackage[normalem]{ulem}
\usepackage[
  citecolor=blue,
  colorlinks,
  linkcolor=blue,
  urlcolor=blue,
]{hyperref}

\begin{document}
\title{Occasional uncoupling overcomes measure desynchronization}
\author{Anupam Ghosh}
\email{anupamgh@iitk.ac.in}
\affiliation{
	Department of Physics,
	Indian Institute of Technology Kanpur,
	Uttar Pradesh 208016, India
}\author{Tirth Shah}
\email{tirth.shah@fau.de}
\affiliation{
	Max Planck Institute for the Science of Light,
	Staudtstraße 2, Erlangen 91058, Germany
}
\affiliation{
	Department of Physics,
	University of Erlangen-Nürnberg,
	Staudtstraße 7, Erlangen 91058, Germany
}
\author{Sagar Chakraborty}
\email{sagarc@iitk.ac.in}
\affiliation{
	Department of Physics,
	Indian Institute of Technology Kanpur,
	Uttar Pradesh 208016, India
}
%
\begin{abstract}
Owing to the absence of the phase space attractors in the Hamiltonian dynamical systems, the concept of the identical synchronization between the dissipative systems is inapplicable to the Hamiltonian systems for which, thus, one defines a related generalized phenomenon known as the measure synchronization. A coupled pair of Hamiltonian systems---the full coupled system also being Hamiltonian---can possibly be in two types of measure synchronized states: quasiperiodic and chaotic. In this paper, we take representative systems belonging to each such class of the coupled systems and highlight that, as the coupling strengths are varied, there may exist intervals in the ranges of the coupling parameters at which the systems are measure desynchronized. Subsequently, we illustrate that as a coupled system evolves in time, occasionally switching off the coupling when the system is in the measure desynchronized state can bring the system back in measure synchrony. Furthermore, for the case of the occasional uncoupling being employed periodically and the corresponding time-period being small, we analytically find the values of the on-fraction of the time-period during which measure synchronization is effected on the corresponding desynchronized state.
 \end{abstract}
\maketitle

\begin{quotation}
As the most simple example, consider two identical Hamiltonian (sub)systems---each of one degree of freedom---coupled in such a fashion that the resultant two degree of freedom system is also a Hamiltonian system, and its phase space trajectory is either quasiperiodic or chaotic. Further consider the situation where at a given coupling strength, an orbit of one subsystem, in the long run, passes arbitrarily close to every point visited by an orbit of the other subsystem such that both the orbits lie in the same domain of the phase plane. We term such a Hamiltonian system to be measure synchronized. One general feature of the systems showing measure synchronization is that there may exist a range of values for the coupling strength parameter at which the two subsystems occupy different domains. Such ranges are aptly called measure desynchronization windows. In this paper, we numerically bring forth, and subsequently analytically explain, a counterintuitive phenomenon where occasionally \emph{uncoupling} the subsystems makes the system inside a desynchronization window measure synchronized.   
\end{quotation}
\section{Introduction}
\label{sec:intro}
The ubiquitous phenomenon of synchronization was first scientifically reported about $350$ years ago by Huygens~\cite{huygens73}. Since then synchronization has been scientifically reported in systems of various sizes, say, from the metabolic processes in our cells~\cite{shliom2014} to the extended ecological systems~\cite{bla1999}. Once rather counterintuitive synchronization of the chaotic systems was observed~\cite{yam1983, afr86,pc1990, pec1991}, a new dimension got added into the research in the field of synchronization.
Today, many different kinds of synchronization~\cite{bocc02, bjps2009, pc15} between the chaotic systems are known. An even more counterintuitive result is that sometimes occasionally uncoupling~\cite{stojanovski97,zochowski00,fortuna03, li05, cqh09,sch2015,sch2016,tandon16, buscarino17, ghosh18} two chaotic systems synchronizes them even though they are not synchronous when coupling is active continuously. Naturally, such occasional uncoupling schemes of synchronization are well-received because, among other reasons, they involve transmission of relatively smaller amount of signal among the systems (hence energy cost is lower) and these schemes give robust stable synchronized states for a wider range of coupling parameters.
In a typical case of the chaotic synchronization, two chaotic orbits---one from each of the identical (dissipative) chaotic (sub)systems which are coupled---asymptotically approach each other even if they start from any arbitrary initial phase points in the same basin of attraction. Naturally, thus, absence of any attractor in a Hamiltonian system makes it impossible to realize such synchronization between two Hamiltonian chaotic systems. Intriguingly, a generalization of the identical synchronization has been proposed so as to synchronize two Hamiltonian (sub)systems as well: this particular type of synchronization is called measure synchronization~\cite{hamp99}, where the two orbits---one from each of the identical coupled subsystems---have identical invariant measures~\cite{Eckmann85} on the portion of the phase space that they share. The measure synchronization in Hamiltonian systems has been observed both for the quasiperiodic and the chaotic motions~\cite{wang03}. Moreover, when more than two systems are coupled, one may also witness partial measure synchronization~\cite{wang02, vincent05}, where a proper subset of the coupled subsystems come together into a measure synchronized state.
What interests us in this paper is measure desynchronization: for certain values of the coupling parameters, that measure how strongly two subsystems are coupled, two coupled Hamiltonian subsystems may not be measure synchronized. We show that the occasional uncoupling can overcome this desynchronization and make the two coupled subsystems measure synchronized. That an occasional uncoupling scheme can induce synchronization in coupled dissipative chaotic systems, doesn't necessarily make it obvious that so should be the case when the scheme is employed on the measure desynchronized systems. This is because the successful implementation of an occasional uncoupling scheme in the dissipative systems is mostly \textit{ad hoc} and the relevant tools of analysis, such as, conditional Lyapunov exponents and eigenvalues of the Jacobian of the corresponding linearized transverse dynamics, are simply not applicable to characterize measure synchronization. Researchers use fundamentally different kind of tools to characterize measure synchronization: the average energy~\cite{wang03}, the variations of phase differences~\cite{hamp99}, the root mean square value of oscillations~\cite{gupta17}, the average interaction energy~\cite{wang03, gupta17}, and the Poincar\'e sections~\cite{tian13}. 

In this paper, we successfully employ the on-off coupling scheme---a deterministic occasional uncoupling scheme---on coupled pairs of Hamiltonian systems that are measure desynchronized. In the on-off coupling scheme the coupling parameter is turned on and off periodically with a preset time-period. For the sake of concreteness, we choose to work with the well-studied $\phi^4$ Hamiltonian system~\cite{wang03} and also with another system that we invent; both the systems exhibit the phenomenon of the measure synchronization.
%

\section{Measure desynchronization}
\label{sec:md}
The measure synchronization transition means a crossover from a measure desynchronized state to a measure synchronized state at a critical value of the coupling parameter.  The synchronized state can either be a quasiperiodic solution of the full coupled system or a chaotic solution. In this paper, since we are interested in bringing measure desynchronized state into measure synchronization, it is convenient to classify the reverse transitions from the measure desynchronized state to the corresponding synchronized state as: (i) quasiperiodic to quasiperiodic desynchronization, and (ii) chaotic to quasiperiodic desynchronization. This is better explained with the help of the concrete examples given in the next two subsections.

\subsection{Quasiperiodic to quasiperiodic desynchronization}
\label{coupled_phi4_MD}
\begin{figure}[h]
	\includegraphics[width= 7.6 cm,height= 11 cm, keepaspectratio]{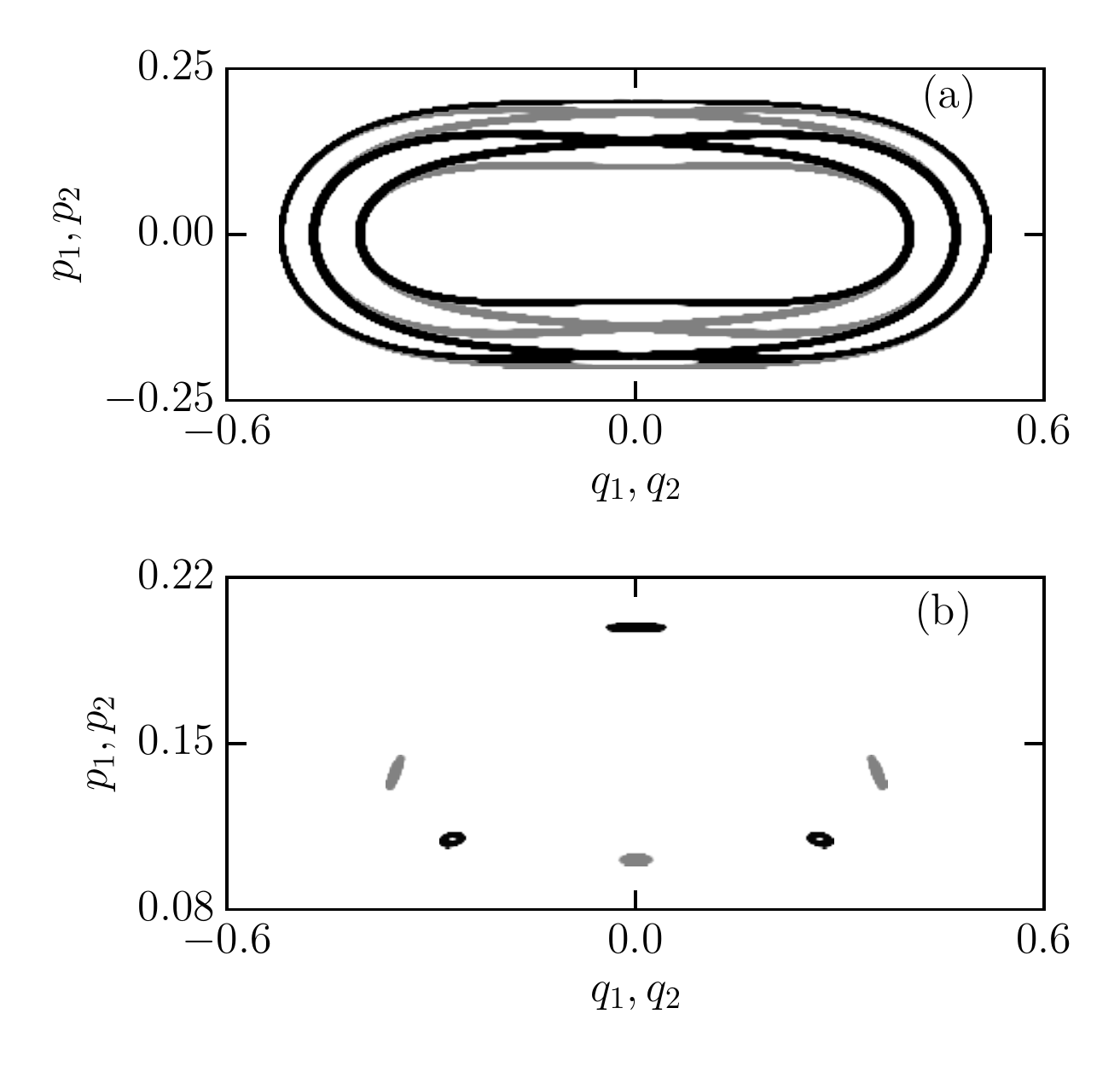}
	\caption{\textcolor{black}{Measure desynchronization: For the QQ-system with initial condition, $(q_1(0), p_1(0), q_2(0), p_2(0)) \equiv (0, 0.1, 0, 0.2)$, the nonidentical (a) phase portraits and (b) Poincar\'e sections highlight the existence of desynchronized state at $K_{\rm QQ} = 0.014$. Black and grey colours respectively refer to $q_1$-$p_1$ and $q_2$-$p_2$ phase spaces.}}
	\label{fig:1}
\end{figure}

Consider an example of a non-integrable bidirectionally coupled system, viz., $\phi ^4$-system~\cite{wang03}, as described by the following Hamiltonian:
\begin{eqnarray}
H\left(q_1,q_2,p_1,p_2\right)&=&\frac{p_1^2}{2}+\frac{q_1^4}{4}+\frac{p_2^2}{2}+\frac{q_2^4}{4}+K_{\rm QQ} \left(q_1-q_2\right)^2\quad \nonumber\\&=&H_1(q_1,p_1)+H_2(q_2,p_2)+H_{\rm coupling}.\label{phi4_Hamiltonian}
\end{eqnarray}
Here, $K_{\rm QQ}$, the coupling strength parameter, is taken to be a real non-negative number. We should view this system as describing the coupling (provided by $H_{\rm coupling}$) between two one degree-of-freedom subsystems: $H_i(q_i,p_i)$; $i=1,2$. The corresponding canonical equations of motion are:
\begin{subequations}
	\begin{eqnarray}
	\dot{q}_1&=&p_1, \\
	\dot{q}_2&=&p_2, \\
	\dot{p}_1&=&-q_1^3+2K_{\rm QQ} \left(q_2-q_1\right), \\
	\dot{p}_2&=&-q_2^3+2K_{\rm QQ} \left(q_1-q_2\right).
	\end{eqnarray}\label{EOM_phi4}
\end{subequations}
If we take the initial condition $(q_1(0), q_2(0), p_1(0), p_2(0))\equiv (0.0, 0.0, 0.1, 0.2)$ with $H(q_1, q_2, p_1, p_2) = 0.025$~\cite{wang03}, desynchronization is observed for $K_{\rm QQ} \in \left[0.0139,0.0145\right]$---we term this closed interval desynchronization window. A desynchronized state in this window is depicted in Fig.~\ref{fig:1}(a) where we can respectively see that the individual  projected phase space plots of the two subsystems at $K_{\rm QQ} = 0.014$. Obviously, the two subsystems occupy different domains of the two dimensional phase space. Note that the subscript `QQ' has been purposefully chosen to remind us that in this case we are dealing with a system trajectory that is quasiperiodic both before and after the measure synchronization transition.  \textcolor{black}{Unless otherwise specified, in this paper, we exclusively work with the aforementioned initial condition for the analysis of the QQ-system without any loss of generality of our results. Later, in the last section of the paper, we elaborately discuss this choice of initial condition.}

\textcolor{black}{Furthermore, the study of Poincar\'e sections (Fig.~\ref{fig:1}(b)) provide an additional support to our conclusions obtained using the projected phase space plots. Construction of a Poincar\'e section\cite{lichtenberg92} and subsequent study of the phase trajectories intersecting it is a convenient technique for studying the dynamics of a Hamiltonian system with more than one degree of freedom. If $N$ is the number degrees of freedom of an autonomous Hamiltonian system, then the corresponding constant energy manifold is a hypersurface of dimension $2N-1$ embedded in the $2N$ dimensional phase space. As far as this paper is concerned, we are interested in $N=2$. Since the Hamiltonian, $H$, is the constant of motion, we can write any variable (say $p_2$) as a function of other three variables (say $q_1, \, q_2, \, \text{and} \, p_1 $). Specifically, from Eq.~(\ref{phi4_Hamiltonian}), it follows that 
	\begin{equation}
	p_2 = \pm \sqrt{2H - p_1^{2} - \frac{q_1^4}{2} - \frac{q_2^4}{2} - 2K_{\rm QQ}(q_1 -q_2)^2}\,.
	\label{p2}
	\end{equation}
Note that if we further restrict ourselves on a plane---the intersection of the constant energy manifold and the manifold given by $q_2 = \text{constant}$---then the dynamics in that plane can be fully specified by using only two variables: $q_1$ and $p_1$.  Of course, the plane $q_2 = \text{constant}$ must be chosen carefully so that there are enough points of intersection of a trajectory with the plane. Also, we note that for a given set of $q_1, \, q_2, \, \text{and} \, p_1 $, we get two values of $p_2$---one positive and the other one negative---each having same magnitude. Thus, for the numerical implementation, it is sufficient to enforce either $p_2>0$ or $p_2<0$ in order to avoid half of the points of intersection that give no new useful information. Additionally,  $q_2 = \text{constant}$ plane has to be given a small width $\varepsilon$ in case one wants to realistically collect enough points of intersection within the limitation of available computational precision and resources. In Fig.~\ref{fig:1}(b), the Poincar\'e sections---$q_1$-$p_1$ and $q_2$-$p_2$ planes---have been plotted for $H(q_1, q_2, p_1, p_2) = 0.025$, $q_2=0$ ($\varepsilon=0.001$), and $p_2 > 0$; and $H(q_1, q_2, p_1, p_2) = 0.025$, $q_1 = 0$ ($\varepsilon=0.001$), and $p_1 > 0$ respectively.}

\subsection{Chaotic to quasiperiodic desynchronization}
\label{example_MD}
In order to describe a system where there is a trajectory that is quasiperiodic before the measure synchronization transition but becomes chaotic after the transition, we have constructed a non-integrable Hamiltonian system:

\begin{eqnarray}
H\left(\theta _1, \theta _2, I_1, I_2\right)&=&\frac{I_1^2}{2}+\frac{I_2^2}{2}-\frac{K_{\rm CQ}}{2}\times \qquad\nonumber\\
&&\left[\cos \left(\theta _1-3\theta _2\right) +\cos \left(3\theta _1-\theta _2\right)\right]\nonumber\\
&=&H_1(I_1)+H_2(I_2)+H_{\rm coupling},\quad\,
\end{eqnarray}
where $H_{\rm coupling}$ is functionally different from the one used in the immediately preceding subsection. Again, the non-negative real $K_{\rm CQ}$ is the coupling strength parameter and the corresponding canonical equations of motion are:
\begin{subequations}
	\begin{eqnarray}
	\dot{\theta}_1&=&I_1, \\
	\dot{\theta}_2&=&I_2, \\
	\dot{I}_1 &=&-\frac{K_{\rm CQ}}{2}\left[\sin \left(\theta _1-3\theta _2\right)+3\sin \left(3\theta _1-\theta _2\right)\right], \\
	\dot{I}_2 &=& \frac{K_{\rm CQ}}{2}\left[3\sin \left(\theta _1-3\theta _2\right)+\sin \left(3\theta _1-\theta _2\right)\right].
	\end{eqnarray}\label{EOM_example}
\end{subequations}
\begin{figure}[h]
	\includegraphics[width= 7.6 cm,height= 11 cm, keepaspectratio]{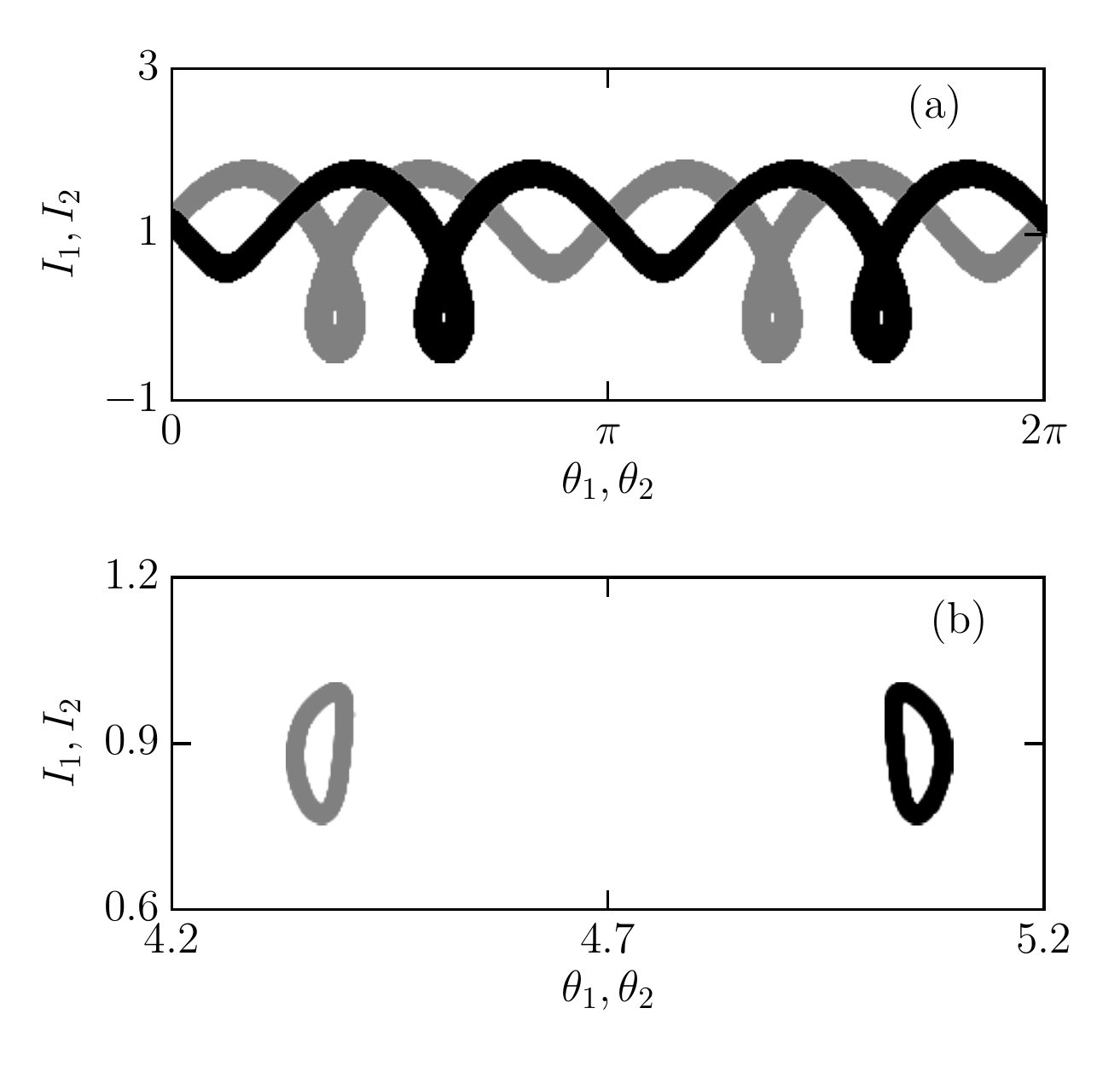}
	\caption{\textcolor{black}{Measure desynchronization: For the CQ-system with initial condition, $(\theta_1(0), I_1(0), \theta_2(0), I_2(0)) \equiv (4.39679, 0.975717, \pi/2, 1.58675)$, the nonidentical (a) phase portraits and (b) Poincar\'e sections highlight the existence of desynchronized state at $K_{\rm CQ} = 1.95$. Black and grey colours respectively refer to $\theta_1$-$I_1$ and $\theta_2$-$I_2$ phase spaces.}}
	\label{fig:2}
\end{figure}
For the initial condition $(\theta _1(0), \theta _2(0), I_1(0), I_2(0)) \equiv (4.39679, \pi/2, 0.975717, 1.58675)$ with $H(\theta_1, \theta_2, I_1, I_2) = 0.2$, a desynchronization window is observed for $K_{\rm CQ} \in \left[1.92, 2.01\right]$. \textcolor{black}{At $K_{\rm CQ}=1.95$, a value inside the window, the desynchronized state is validated by the non-overlapping phase space plots [Fig.~\ref{fig:2}(a)] and the nonidentical Poincar\'e sections [Fig.~\ref{fig:2}(b)] for the two subsystems. For plotting the Poincar\'e sections, we have taken the positive values of the actions, set $H=0.2$, and fixed $\varepsilon=0.001$ while working with the planes $\theta_i=\pi/2$ ($i=1,2$).} \textcolor{black}{Similar to what has been done with the QQ-system, we stick with the above mentioned illustrative initial condition for further analysis of the CQ-system.}

We note that the subscript `CQ' has been chosen to indicate that in this case we are dealing with a system trajectory that is chaotic in the measure synchronized state but becomes quasiperiodic when (reverse) transitions into a desynchronized state. \emph{For later convenience, henceforth, we call this system the CQ-system and $\phi^4$-system as the QQ-system.}
%

%
\section{Detection of measure synchronization}
\label{mms}
Howsoever tempting it might look to detect measure synchronized states by mere inspection of the phase portraits, whether a state is measure synchronized or not can only be established quantitatively. In this section, we discuss the proper quantifications that allow us to detect measure desynchronized windows as is need for the investigation undertaken in this paper.
\subsection{Energy based methods}
In order for the two subsystems to be measure synchronized, the joint phase space probabilities of the generalized coordinates and the generalized momenta for both the subsystems must be identical. Consequently, it is expected that for a measure synchronized state, the time average of any function of the generalized coordinates and the generalized momenta of the individual subsystems should be equal. Thus, the difference of average bare energies ($\Delta E$), 
\begin{equation}
\Delta E = \frac{1}{T_f}\int_{0}^{T_f} \left(H_1-H_2\right) dt,
\end{equation}
should be very close to zero in a measure synchronized state~\cite{wang03}. Here $T_f$ is the large final time till when the system has been evolved. Consequently, in a plot of $\Delta E$ vs the coupling strength parameter, a desynchronized state among the synchronized states could be concluded from a non-zero value of $\Delta E$.
\begin{figure}[h]
	\includegraphics[width= 8.6 cm,height= 12 cm, keepaspectratio]{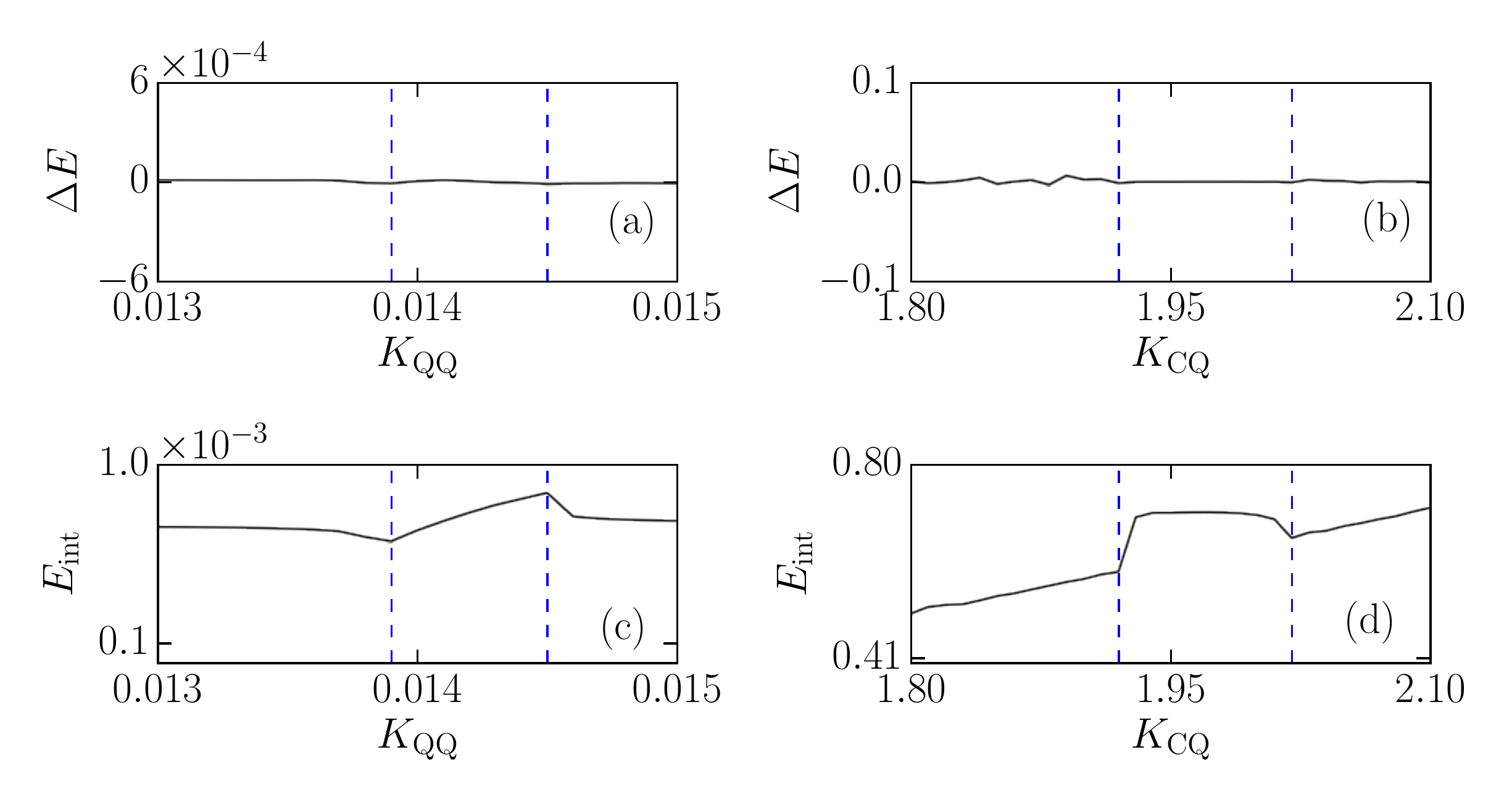}
	\caption{\textcolor{black}{Measure desynchronization windows. Sudden jumps bounding a raised plateau in the plots of (c) $E_{\rm int}$ vs $K_{\rm QQ}$ and (d) $E_{\rm int}$ vs $K_{\rm CQ}$ show the presence of the desynchronized windows for the QQ-system and the CQ-system respectively. Observe, however, the fallacy presented by the plots (a) $\Delta E$ vs $K_{\rm QQ}$ and (b) $\Delta E$ vs $K_{\rm CQ}$: Although there are desynchronization windows, there are no appreciable non-zero shift in the values of $\Delta E$.}}
	\label{fig:3}
\end{figure}
However, contrary to what can be concluded from the phase plots, since there is no non-zero fluctuation in the value of $\Delta E$ plotted against $K_{\rm QQ}$ [Fig.~\ref{fig:3}(a)], it seems to (falsely) indicate the absence of the desynchronization window in the QQ-system. Similar is the case with the CQ-system: in Fig.~\ref{fig:3}(b), $\Delta E$  erroneously does not indicate any desynchronization window when plotted against $K_{\rm CQ}$. This problem with the usage of $\Delta E$ to detect the measure synchronized states is also seen in other types of coupling~\cite{gupta17}. In passing, we remark that the order of $\Delta E$ in Fig.~\ref{fig:3}(b) is much larger than that of in Fig.~\ref{fig:3}(a). This is because for a chaotic trajectory, getting $\Delta E \rightarrow 0$ requires us to evolve the system for much larger time (i.e., $T_f$ should be very large). 

In the similar spirit, another quantity called the average interaction energy~\cite{wang03}, 
\begin{equation}
E_{\rm int} = \frac{1}{T_f}\left|\int_{0}^{T_f} (H_{\rm coupling}) dt\right|,
\end{equation}
may also be used to detect the measure desynchronized state.
In Fig.~\ref{fig:3}(c), kinks or sudden jumps bounding a raised or a lowered plateau in the plot of $E_{\rm int}$ \textcolor{black}{vs} $K_{\rm QQ}$ mark the boundaries of the desynchronization window for the QQ-system. Same is the case with the CQ-system as seen in Fig.~\ref{fig:3}(d). 

\subsection{Joint probability densities}
\textcolor{black}{Due to the apparent contradictory results presented by the two methods discussed above, we feel that it is best to go back to the first principles to quantitatively confirm the conclusions rendered by the energy based methods. Thus, all we want to check is whether the two orbits---one from each of the identical coupled subsystems---have identical invariant measures on the portion of the phase space that they share. One way of checking it is to find the joint probability distributions of the two sets of the phase points covered by the two orbits of the two subsystems and show that the distributions are identical implying measure synchronization. For practical reasons, we consider the two joint distributions to be identical when the corresponding values of the joint probability density functions in each of the small bins of same size are equal up to a small additive constant, $\tau$. If the difference in the values of the two probability densities in any bin is more than the threshold value, $\tau$, we define the system to be in measure desynchronized state. We implement this idea quantitatively as follows:}
\begin{figure*}[t]
	\includegraphics[width= 17.6 cm,height= 22 cm, keepaspectratio]{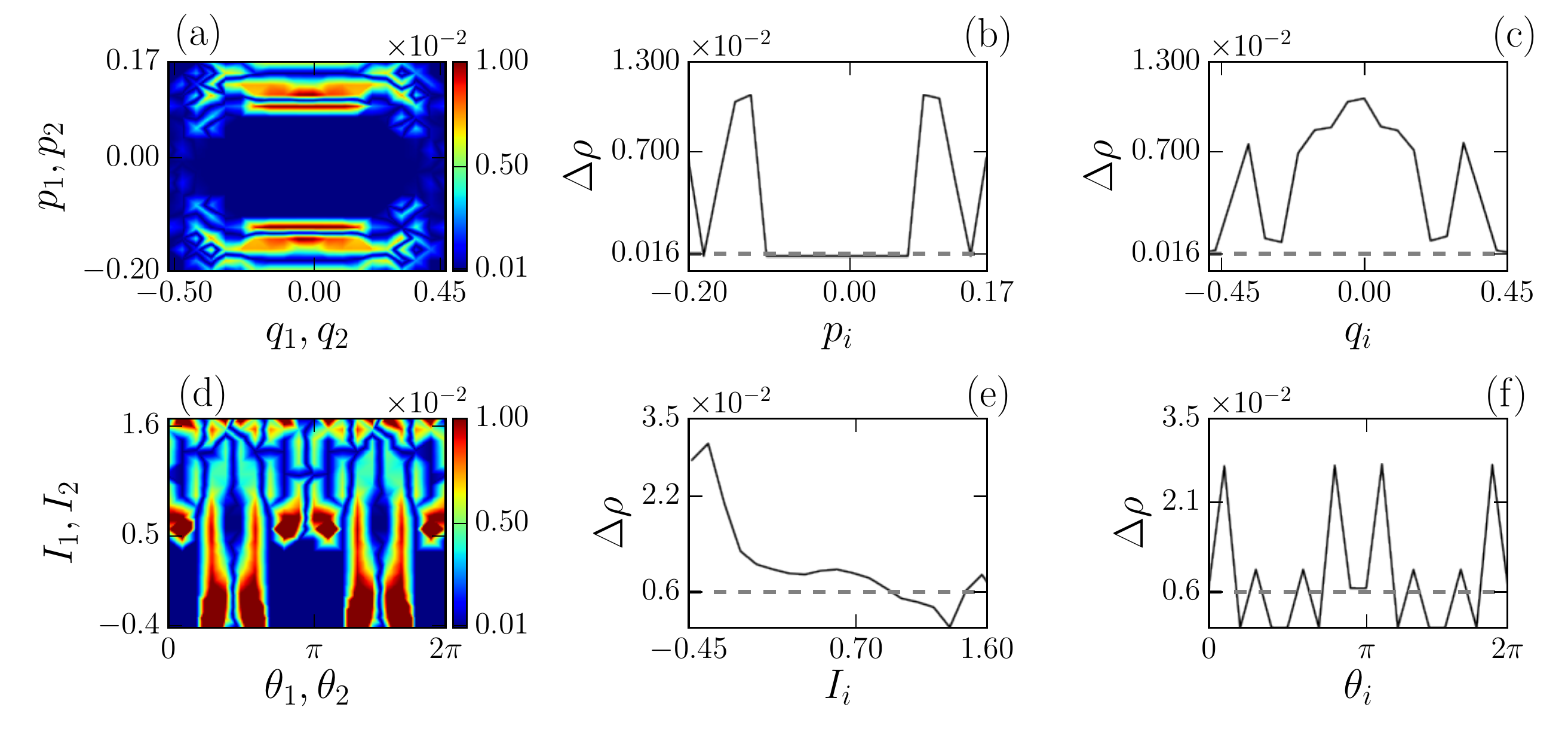}
	\caption{\emph{(color online)} \textcolor{black}{Measure desynchronized states. The absolute differences, $\Delta\rho$, of the joint probability densities have been depicted for the QQ-system and the CQ-system with $K_{\rm QQ} = 0.014$ and $K_{\rm CQ} = 1.95$ respectively. Subplots (a), (b), and (c) refer to continuous coupling in the QQ-system, while subplots (d), (e), and (f) correspond to continuous coupling in the CQ-system. The color-bars in subplots (a) and (d) quantify the magnitude of $\Delta\rho$. Subplot (b) illustrates how $\Delta \rho$ varies with $p_i$ ($i=1,2$) at an illustrative value for coordinate $q_1= q_2 = 0$. Similarly,  subplot (c) illustrates how $\Delta \rho$ varies with $q_i$ ($i=1,2$) at an illustrative value for momenta $p_1= p_2 = 0.12$. The gray dashed lines in (b) and (c) correspond to the threshold $\tau=1.6\times 10^{-4}$. Similarly, subplots (e) and (f) respectively demonstrate the variation of $\Delta \rho$ with the actions keeping the angles fixed and with the angles keeping the actions constant. The grey dashed lines in (e) and (f) corresponds to the threshold $\tau=6\times 10^{-3}$.}}
	\label{fig:4}
\end{figure*}
Consider either the QQ-system or the CQ-system and a trajectory $(q_1(t),q_2(t),p_1(t),p_2(t))$ that is a continuous sequence of phase points in the four-dimensional phase space. First we find 
\begin{eqnarray*}
&&q_{\rm min}:=\min(\min(q_1(t)),\min(q_2(t))),\\
 &&q_{\rm max}:=\max(\max(q_1(t)),\max(q_2(t))),\\ 
 &&p_{\rm min}:=\min(\min(p_1(t)),\min(p_2(t))),\\ 
 &&p_{\rm max}:=\max(\max(p_1(t)),\max(p_2(t))).
\end{eqnarray*}
Here, $p$'s and $q$'s denote the conjugate momenta and the generalized coordinates respectively; they can also mean the action-angle variables used in the CQ-system. Subsequently, we divide the ranges $(q_{\rm min},q_{\rm max})$ and $(p_{\rm min},p_{\rm max})$ in equal number of bins, say $M$, such that the square cells of area $\Delta q\Delta p$ are created, where $\Delta q:=(q_{\rm max}-q_{\rm min})/M$ and $\Delta p:=(p_{\rm max}-p_{\rm min})/M$. Thus, if $\rho_i(q_i,p_i)$ ($i=1,2$) is the joint probability density function, then $\rho_i(q_i,p_i)\Delta q\Delta p$ gives the fraction of points lying in the square cell centered at $(q_i,p_i)$. In this paper, we characterize a system to be in measure synchrony if  $\Delta\rho:=|\rho_1 (q, p)-\rho_2 (q, p)|\le\tau$ $\forall\,(q,p)$ where $\tau$ is a predefined threshold number taken to be reasonably small.  

As an illustration, Fig.~\ref{fig:4}(a) exhibits $\Delta\rho$, with $M$ set to $20$ for the QQ-system for the continuous coupling.  For $M=20$, we get $400$ square cells at each of which the condition, $\Delta\rho<\tau=1.6\times10^{-4}$, ensures measure synchronization when the on-off coupling is in action. The CQ-system has been studied similarly [Fig.~\ref{fig:4}(d)], where we have taken $M = 20$ and the corresponding threshold $\tau=6\times10^{-3}$. \textcolor{black}{While much lower values of $M$ would be useless as there won't be enough square  cells to capture the local dynamics, any higher value of $M$ ($\gtrsim20$) doesn't change our conclusions in this paper; however, one may have to evolve the systems for much longer time so that there are enough points in each square cell to capture the nature of the local dynamics. In fact, if we evolve the systems for much longer time, we could work with much lower values of $\tau$. Thus, one needs to strike a balance between choosing as small a value of $\tau$ possible and running the numerical codes for long time. Also, it is not unexpected that $\tau$ would be different for different systems evolved for same time: practical choice of $\tau$ depends on the fractions of the phase space that the phase trajectories covers in the time the corresponding system evolves}. We emphasize that our conclusions in this draft are not based only on the nature of $\Delta\rho$, but also on the nature of the the plots of $E_\text{int}$.

In rest of the paper, we adopt all the three quantitative tools---$\Delta E$, $E_{\rm int}$, and  joint probability distributions---of finding the desynchronization windows. As we have already discussed that $\Delta E$ is inconclusive in detecting the measure desynchronization, the study of $\Delta E$ has been done mostly for the sake of completeness.
\section{Occasional uncoupling}
\label{sec:ds}
After understanding the measure synchronization and the measure desynchronization for the Hamiltonian systems, an immediate natural curiosity would be whether other concepts and phenomena related to the synchronization of the dissipative systems can also be extended to encompass measure synchronization. In this context, we wonder if the measure desynchonized state can be brought back to synchrony without changing the coupling strength. As discussed in the introduction to this paper, the occasional coupling schemes~\cite{stojanovski97,zochowski00,fortuna03, li05, cqh09,sch2015,sch2016,tandon16, buscarino17} are known to be successful in inducing synchronization in the coupled chaotic dissipative systems when they are not in synchrony if continuously coupled. One such typical scheme is the on-off coupling scheme~\cite{cqh09}. The on-off coupling periodically switches the coupling between the subsystems on and off. Defining $T$ and $\theta$ ($\theta\in[0,1]$) be the on-off period and the on-fraction respectively, the coupling is active when the time $t$ is such that $nT \le t < (n+\theta)T$ ($n\in\{0,1,2,\cdots\}$) and is inactive if $(n + \theta)T \le t < (n+1)T$. 
\begin{figure*}[t]
	\hspace{-9.6 mm}
	\includegraphics[scale = 0.4]{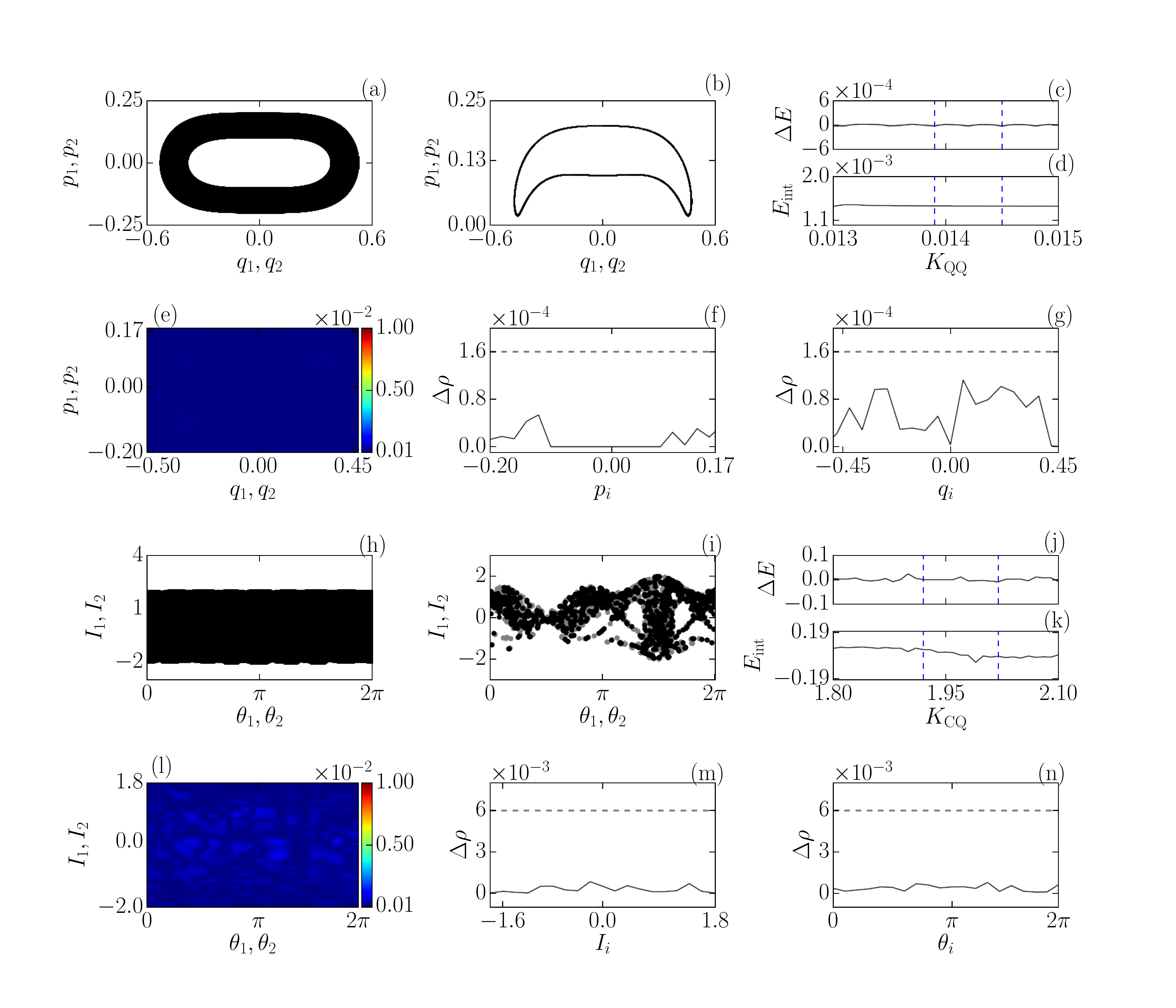} 
	
	\caption{ \textit{(Color online)} \textcolor{black}{The on-off coupling overcomes measure desynchronization. We choose $(T, \theta)= (0.3, 0.5)$ and $(T, \theta) = (0.05, 0.5)$ for the QQ-system and the CQ-system respectively. The subplots in the top panel exhibit (a) the overlapping phase trajectories, (b) the overlapping Poincar\'e sections, (c) the vanishing of $\Delta E$, and (d) an $E_{\rm int}$ vs $K_{\rm QQ}$ curve without any abrupt jumps. These imply the suppression of the desynchronization window in the QQ-system due to the occasional uncoupling. Similarly, plots (h)--(k), depict the supression of the desynchronization window in the CQ-system. In subplots (e) and (l), the absolute difference, $\Delta\rho$, of the joint probability densities have been shown for the QQ-system and the CQ-system respectively.  Here, the color-bars indicate the values of $\Delta\rho$; $\Delta\rho\approx0$ indicate measure synchronization. Subplot (f) illustrates how $\Delta \rho$ varies with $p_i$ ($i=1,2$) at $q_1= q_2 = 0$, and similarly,  subplot (g) illustrates how $\Delta \rho$ varies with $q_i$ ($i=1,2$) at $p_1= p_2 = 0.12$. The grey dashed lines in (f) and (g) correspond to the threshold $\tau = 1.6\times 10^{-4}$. We note that in subplots (e), (f), and (g), the maximum value of $\Delta \rho$ lower than the threshold, thus, showcasing the effect of the measure synchronization induced by the on-off coupling. This is also the case for the CQ-system as exhibited in subplots (l), (m), and (n)---analogues of (e), (f), and (g)---where the corresponding threshold is $\tau = 6\times 10^{-3}$. We mention that, except in the subplots (d) and (k), we have chosen illustrative values $0.014$ and $1.95$ inside the corresponding desynchronization windows for $K_{\rm QQ}$ and $K_{\rm CQ}$ respectively.}}
	\label{fig:5}
\end{figure*}

We must appreciate that it is not at all clear \emph{a priori} if the success of the on-off coupling in dissipative systems must carry over to the case of measure desynchronization. The mechanism behind the success of the on-off coupling scheme or such similar occasional coupling schemes (like the transient uncoupling scheme~\cite{sch2015}) in dissipative system may be traced to the favourable set of spectra of eigenvalues of the Jacobians~\cite{tandon16} found at each point of the response subsystem's trajectory. (Instead of these eigenvalues, one could  have used other quantities like local Lyapunov exponent~\cite{abr92,eck1993} or  eigenvalues of the symmetrized Jacobian~\cite{doerner91,ghosh18}). Also, in effect, the negativity of the maximum conditional Lyapunov exponent is the necessary condition for establishing synchronized state~\cite{pea1997,pecora98}. However, due to the absence of the phase space attractors, the measure synchronization of Hamiltonian systems can not be characterized or explained using the aforementioned quantities. Moreover, the measure synchronization may be observed for both the quasiperiodic and the chaotic trajectories. It is thus crystal clear that whether on-off coupling is going to be successful in bringing about measure synchronization is an open interesting question.
For further discussion and to explicitly spell out the implementation of the on-off coupling scheme, we mathematically represent the two Hamiltonian subsystems bidirectionally coupled with coupling strength parameter $K$ as follows:
\begin{eqnarray}
\label{vector_field}	
\dot{\textbf{x}}  &=& \textbf{f}\left(\textbf{x}\right)+\tilde{\chi}_{T,\theta}(t){K}  {\textsf{C}} \cdot \mathbf{g}\left(\textbf{x}\right),
\end{eqnarray} 
where 
\begin{equation}
\tilde {\chi}_{T,\theta}(t) := \begin{cases}
1  \text{ for } nT \le t < (n+\theta)T,\\
0  \text{ for }(n+\theta)T \le t <(n+1)T.
\end{cases}
\end{equation}
For the QQ-system, the $4$ dimensional column vectors $\textbf{x}$, $\textbf{f}(\textbf{x})$, and $\textbf{g}(\textbf{x})$ are respectively equal to $\left(q_1,q_2,p_1,p_2\right)$, $(p_1,p_2,-q_1^3,-q_2^3)$, and $(0,0,2 q_2 - 2 q_1, -2 q_2 + 2 q_1)$. Also, $K=K_{\rm QQ}$ and the $4\times4$ coupling matrix $\textsf{C}_{ij}=\delta_{i3}\delta_{j3}+\delta_{i4}\delta_{j4}$ ($\delta$ being the Kronecker delta). Similarly, for the CQ-system, $\textbf{x}=\left(\theta_1,\theta_2,I_1,I_2\right)$, $\textbf{f}(\textbf{x})=(I_1,I_2,0,0)$, $K=K_{\rm CQ}$, $\textsf{C}_{ij}=\delta_{i3}\delta_{j3}+\delta_{i4}\delta_{j4}$, and $\textbf{g}(\textbf{x})=(0,0,- \frac{1}{2}\sin(\theta_1 - 3\theta_2) -\frac{3}{2}\sin(3\theta_1 - \theta_2), \frac{3}{2} \sin(\theta_1 - 3\theta_2) + \frac{1}{2}\sin(3\theta_1 - \theta_2))$.
\subsection{Overcoming measure desynchronization}
\label{sec:ds}
%

Now, for the QQ-system, we choose $T=0.3$ and $\theta=0.5$ by trial-and-error to impart synchronization inside the desynchronization window, $0.0139\le K_{\rm QQ}\le0.0145$, at $K_{\rm QQ}=0.0140$. \textcolor{black}{As has been the case for the corresponding continuous coupling [Fig.~\ref{fig:1}(b)], in Fig.~\ref{fig:5}(b), the Poincar\'e sections---$q_1$-$p_1$ and $q_2$-$p_2$ planes---have been plotted for $H(q_1, q_2, p_1, p_2) = 0.025$, $q_2=0$ ($\varepsilon=0.001$), and $p_2 > 0$; and $H(q_1, q_2, p_1, p_2) = 0.025$, $q_1 = 0$ ($\varepsilon=0.001$), and $p_1 > 0$ respectively. Fig.~\ref{fig:5}(a)--(d) show how at $K_{\rm QQ}=0.0140$, the on-off coupling leads to (i) the overlapping phase trajectories, (ii) the overlapping Poincar\'e sections, (iii) the vanishing of $\Delta E$, and (iv) an $E_{\rm int}$ vs $K_{QQ}$ curve without any abrupt jumps corresponding to the boundaries of the desynchronization window. This means that the on-off coupling has brought synchrony to the otherwise desynchronized state. We have confirmed this conclusion by establishing in Fig.~\ref{fig:5}(e)--(g) that the joint probability distributions of the two subsystems are equal within the threshold $\tau=1.6 \times 10^{-4}$.}

Again, as far as the CQ-system is concerned, we take $T=0.05$ and $\theta=0.5$ to induce the measure synchronization at $K_{\rm CQ}=1.95$ inside the desynchronization window: $1.92\le K_{\rm CQ}\le2.01$. \textcolor{black}{The Poincar\'e sections in Fig.~\ref{fig:5}(i) for the CQ-system are plotted with the positive values of the actions, $H=0.2$, and $\varepsilon=0.001$ while working with the planes $\theta_i=\pi/2$ ($i=1,2$). These conditions are exactly same as has been chosen for the corresponding continuous case, thereby validating the fact that the synchronization has been brought about by the occasional uncoupling. We note from Fig.~\ref{fig:5}(h)--(k) that all the characterizations of measure synchronization---the overlapping phase trajectories, the overlapping Poincar\'e sections, the vanishing of $\Delta E$, and the $E_{\rm int}$ vs $K_{CQ}$ curve without any abrupt jumps---validate the success of the on-off coupling in effecting synchronization. Above all, as seen in Fig.~\ref{fig:5}(l)--(n), we have conclusively shown that the joint probability distributions are identical (within the threshold $\tau= 6 \times 10^{-3}$) as expected in a synchronized state.}

\emph{Thus, in conclusion, we can decisively say that the on-off coupling does overcome measure desynchronization.}
%
\subsection{Choosing on-off period and on-fraction}
We have seen that using an appropriate combination of the parameters $T$ and $\theta$, the on-off coupling can induce synchronization in an otherwise measure desynchronized state. As presented, the choice of these parameters appeared to be \emph{ad hoc}. In fact, we have found that the process of choosing a value of  $\theta$ that can impart synchronization, if at all, for a given $T$ is indeed a matter of trial-and-error. Nevertheless, fortunately, if $T$ is small enough compared to the timescale of the system under consideration, then there exists a well-defined prescription for how to choose $\theta$. This is exactly what we intend to discuss in this section.

From Eq.~(\ref{vector_field}), $\textbf{x}(t+T)$ can be written as:
\begin{eqnarray}
\label{vector_field_TE}
\textbf{x}\left(t+T\right)&=&\textbf{x}(t)+\int _t^{t+T}\textbf{f}\left(\textbf{x}(t')\right) dt' \nonumber\\ &&+\int _t^{t+T} \tilde{\chi}_{T,\theta}(t')K\textsf{C} \cdot \mathbf{g}(\textbf{x}(t')) dt'.
\end{eqnarray}
Evidently, if $T$ is so small that the vector functions $\textbf{f}$ and $\textbf{g}$ do not vary substantially, then it follows that
\begin{equation}
\label{approx_vector_field_TE}
\textbf{x}\left(t+T\right) \approx \textbf{x}(t)+\textbf{f}\left(\textbf{x}(t)\right) T +\theta K\textsf{C} \cdot \mathbf{g}(\textbf{x}(t))T,
\end{equation}
where $\textbf{f}$ and $\textbf{g}$ have been assumed to be constant over the time $T$ and, in the last term, we have explicitly incorporated the fact that the coupling is active only over a fraction $\theta$ of $T$. Thus, it is straightforward to conclude by inspection that one can think of the system under the action of the on-off coupling as the system under the action of continuous coupling but with an effective lower value of coupling strength given by:
\begin{equation}
\label{eq:eff}
K_{\text{eff}}=\theta K.
\end{equation}
\begin{figure}[t]
	\includegraphics[width=20.1cm,height=8.0cm, keepaspectratio]{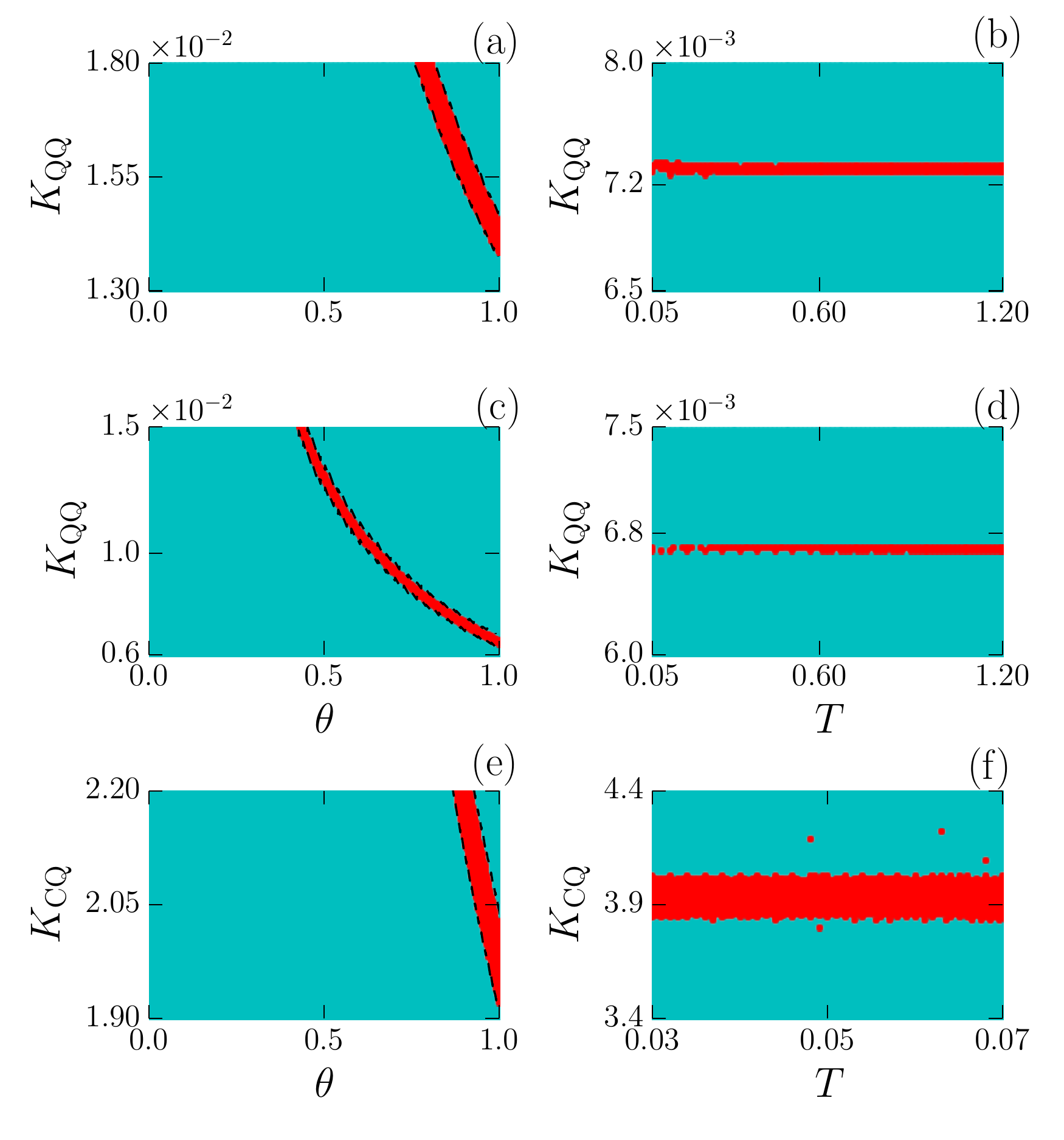}
	\caption{ \textit{(Color online)} Desynchronization windows shift along rectangular hyperbola. Cyan indicates synchronized states and red indicate desynchronized states. In (a) and (c), the two isolated desynchronization windows, $0.0139\le K_{\rm QQ}\le0.0145$ and $6.43 \times 10^{-3}\le K_{\rm QQ} \le 6.56 \times 10^{-3}$, respectively are seen shifting along rectangular hyperbolic paths with decrease in $\theta$ ($T=0.3$). $\theta$ is fixed at 0.5 in (b) and (d). Similarly, in (e) the desynchronization window of the CQ-system is seen following the relationship: $K\propto1/\theta$ ($T=0.05$). $\theta=0.5$ in (f). The black dashed lines in (a), (c), and (e) are the analytically expected hyperbolic curves that bound the shifting windows perfectly.}
	\label{fig:6}
\end{figure}

Armed with this simple and elegant result, we can now find the values of the on-fraction, $\theta$, such that an entire desynchronization window may be gotten rid of by inducing the measure synchronization therein. Only constraint we have to respect is that $T$ should be small compared to the corresponding system's timescale. Consequently, since the approximate system-timescales ($T_s$) are $16$ and $18$ (deduced from the time series of the coordinates) for the QQ-system and the CQ-system, in what follows we conveniently take $T$ as $0.3$ and $0.05$ respectively. It must also be noted that, as long as $T\ll T_s$, Eq.~(\ref{eq:eff}) is independent of $T$.

First, let us focus on the desynchronization window: $0.0139\le  K_{\rm QQ} \le0.0145$ in the QQ-system. \textcolor{black}{Fig.~\ref{fig:6}(a) shows that the desynchronization window shifts toward the higher values of $K_{\rm QQ}$ as $\theta$ decreases}. As the on-off coupling scheme is implemented by decreasing $\theta$, we note that the desynchronization window shifts to the higher values of the coupling strength. It means that with the decrease in the on-fraction, the on-off coupling induces measure synchronization \textcolor{black}{in an increased fraction of the window}. The quantitative manner in which the synchronization is effected is in line with the prediction of Eq.~(\ref{eq:eff}): for the constant values $(K_{\rm QQ})_{\rm eff} \approx 0.0139$ and  $(K_{\rm QQ})_{\rm eff} \approx 0.0145$, we respectively get the left and the right boundaries (dashed black rectangular hyperbolic curves---$\theta K_{\rm QQ}=(K_{\rm QQ})_{\rm eff}$---in Fig.~\ref{fig:6}(a)) of the shifting window (exhibited as red band in Fig.~\ref{fig:6}(a)). Fig.~\ref{fig:6}(b) validates the result that for a fixed optimal $\theta$, the desynchronization window, and hence the induced measure synchronization, is independent of the small values of $T$. We must clarify that we have presented the aforementioned window in isolation. There are various other desynchronization windows present in the QQ-system, e.g., $6.43 \times 10^{-3}\le  K_{\rm QQ} \le 6.56 \times 10^{-3}$ that is presented in isolation in Fig.~\ref{fig:6}(c), also in isolation, for clarity. We again note that the on-off coupling induced synchronization inside this window is as predicted by Eq.~(\ref{eq:eff}) and the phenomenon is independent of the on-off period as long as it is small enough [see Fig.~\ref{fig:6}(d)].

Further, as graphically elaborated in Fig.~\ref{fig:6}(e)-(f), we have verified that even for the CQ-system, the corresponding desynchronization window shifts to the higher values of the coupling strength as the on-fraction decreases. Each $(\theta, K_{\rm CQ})$ point in the window shifts along an analytically predicted rectangular hyperbola and is independent of the on-off period that is taken to be much smaller than the system's timescale. A few scattered points beyond the desynchronization band in Fig.~\ref{fig:6}(f) are, we believe, due to fact that the chaotic nature of the system demands that we evolve the system to extremely large time to get equal joint probability distributions. A rather artificial way of getting rid of them could be just to take a higher value of the threshold $\tau$. However, we have chosen to present the plot as it is in order to render the readers mindful of such caveats in our study.

In conclusion, we can now give a straightforward answer to the question that which value of $\theta$ is effective in causing measure synchronization of a desynchronized state inside a desynchronized window. Suppose that $T\ll T_s$ and there exists a single desynchronization window, $K_1\le K\le K_2$, of the system. \emph{Then the measure desynchronized state of the system at any $K\in[K_1,K_2]$ can be synchronized by choosing any value of $\theta$ less than $K_1/K$.}
%

\section{Discussion and conclusion}
\label{sec:cd}
\begin{figure}[t]
	\includegraphics[width=20.1cm,height=8.0cm, keepaspectratio]{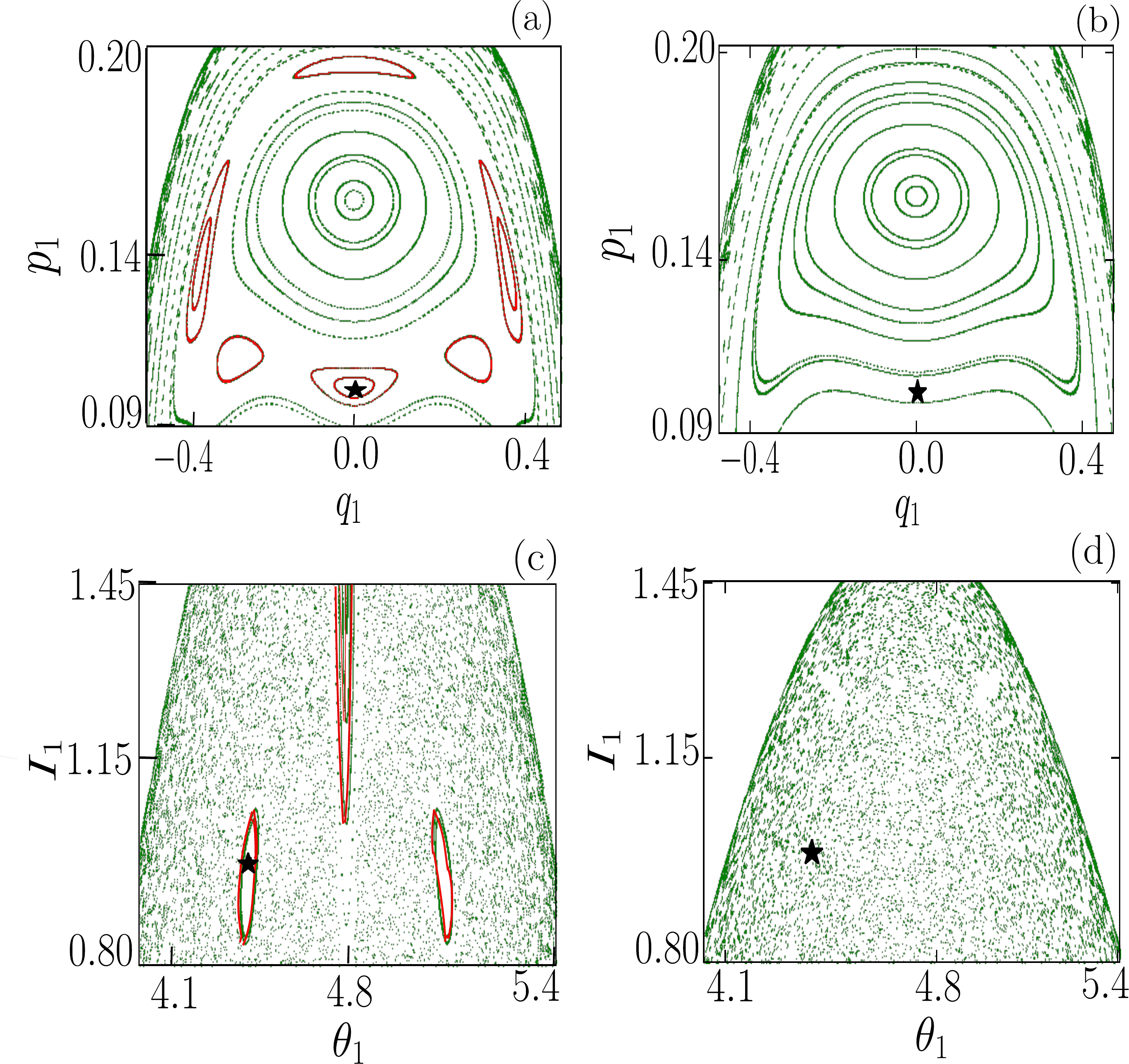}
	\caption{ \emph{(Color online)} \textcolor{black}{The Poincar\'e sections of the first subsystem for the QQ-system [subplots (a) and (b)] and the CQ-systems [subplots (c) and (d)] are plotted using $50$ different set of initial conditions. We have chosen $K_{\rm QQ} = 0.014$  and $K_{\rm CQ} = 1.95$. Subplots (a) and (c) are for the continuously coupled system, while subplots (b) and (d) are for the cases where on-off scheme has been employed such that $(T,\theta)$ is equal to $(0.3,0.5)$ and $(0.05,0.5)$ respectively. Thus, effective $K_{\rm QQ} = 0.014 \times 0.5= 0.007$ and effective $K_{\rm CQ} = 1.95 \times 0.5= 0.975$ respectively (see Eq.~\ref{eq:eff}). The black stars in the plots for the QQ-system and the CQ-system correspond to the initial conditions---$(q_1(0), p_1(0), q_2(0), p_2(0))= (0, 0.1, 0, 0.2)$ and $(\theta_1(0), I_1(0), \theta_2(0), I_2(0)) = (4.39679, 0.975717, \pi/2, 1.58675)$ respectively---used extensively in the paper. If an initial condition in is the red region then a desynchronized state is observed, otherwise a synchronized state is observed.}}
	\label{fig:7}
\end{figure}
\textcolor{black}{Any phenomenon associated with the dynamics in the phase space of a Hamiltonian system is critically dependent on the structure of the phase space and the initial conditions. The measure synchronization is such a phenomenon. This dependence on initial conditions comes from two factors: (i) there cannot be any attractor in a Hamiltonian system, and hence, every orbit is capable of showing asymptotically different dynamical behaviour compared to any other orbit; and (ii) an initial condition fixes the energy of the autonomous Hamiltonian system, implying that the initial condition effectively becomes a parameter of the system unlike what happens in a dissipative system. Thus, of course, if we choose different set of initial conditions~\cite{hamp99,wang03}, the system behaves differently; in fact, mere change in the initial condition may change the dynamics from quasiperiodic to chaotic or vice versa. In passing, we also remark that one can~\cite{wang02, wang03, vincent05, gupta17} chose the initial conditions so as to render the average interaction energy zero and, thus, the initial conditions can be treated as a parameter independent of coupling strength. In any case, the existence of the measure synchronization (and, similarly, a desynchronization window) does and should depend on the initial conditions. However, one must realize that this paper is not at all about finding the measure synchronization or desynchronization windows. There is a plethora of evidences that they exist. Our main aim is to show that when a system is measure desynchronized, it can be brought into a measure synchrony by implementing an occasional coupling scheme. This result is not at all obvious a priori. (Why occasional coupling schemes work for dissipative chaotic systems synchronization is also an open question.)} 

\textcolor{black}{The motivation behind choosing the specific systems investigated in our manuscript is that they are among the simplest possible QQ-system and CQ-system that possess desynchronization windows so that we can test occasional uncoupling schemes on them. As seen in Fig.~\ref{fig:7}(a) and Fig.~\ref{fig:7}(c), in each of the systems for a fixed value of the corresponding coupling parameter, there exists a set of initial conditions (within the red coloured resonance islands in the figures) that lead to some desynchronization states, and eventually to some desynchronized window when the coupling parameter is varied. Although it is not crucial for the main result of this paper, we mention that for obvious reasons, a set of initial conditions from the green coloured regions in these figures leads only to synchronized states and no desynchronization window is seen for the same fixed values of the coupling parameters. This highlights the expected feature of the Hamiltonian systems that the choice of initial conditions for studying the measure synchronization is strongly dependent on the corresponding phase space structure. Intuitively speaking, the occasional switching off of the coupling makes the phase trajectory of one of the subsystems to jump into the region of phase space occupied by the other such that the subsystems' subsequent dynamics are naturally measure synchronized [see Fig.~\ref{fig:7}(b) and Fig.~\ref{fig:7}(d)]. However, the question is: what kind of occasional uncoupling enforces such a helpful jump? As discussed in the paper, when the on-off coupling scheme is implemented, any arbitrary $\theta$ and $T$ are not able to bring back the measure synchronization. Therefore, our conclusion about how to find the right combination of $\theta$ and $T$ to bring about measure synchronization in a desynchronization window is a non-trivial result. \emph{In summary, we have numerically and analytically shown whether, when, and how the on-off coupling scheme induces either quasiperiodic or chaotic measure synchronization in a desynchronized state of a Hamiltonian system.}}

This scheme of effecting measure synchronization without directly or explicitly changing the coupling strength is very robust. To do a quick check, we introduced an additive random noise, $D \zeta(t)$, in both the QQ-system and the CQ-system. Here, $D$ is the noise amplitude and $\zeta(t)$ has a Gaussian random distribution with zero mean, unit variance, and is temporally delta-correlated. On taking the value of $D$ approximately ten times smaller than the other deterministic terms in the equations of motion, we could still effect measure synchronization (within noise fluctuations) in the desynchronization window. Moreover, the phenomenon of the shift of the window along a \textcolor{black}{hyperbola with decreasing} on-fraction also remains intact. In passing, we remark that it could be insightful to analytically analyze how the noise manifests itself, if at all, in modifying~\cite{sam2014} the system parameters---most importantly the coupling strength parameter.

While our study is complete in itself, it does open up some new questions. We recall that our analytical result, validated by the numerical experiments, has been based on the assumption that $T\ll T_s$. It thus is intriguing that the measure synchronization can be imparted on the desynchronized states even with larger values of $T$ and a corresponding appropriate $\theta$, e.g., for the QQ-system, the combination $(T, \theta)=(2, 0.7),\, (6, 0.3),\,\textrm{or}\,(10, 0.7)$ and for the CQ-system, the combination $(T,\theta)=(0.3, 0.7),\,(2, 0.7),\,\textrm{or}\,(15, 0.8)$ are capable of inducing measure synchronization in the otherwise desynchronization states at $K_{\rm QQ}=0.014$ and $K_{\rm CQ}=1.95$ respectively. We do not have an answer to why so happens; a detailed theory explaining it is missing and could be a challenging problem to tackle in future. 

It should be borne in mind that the on-off coupling method of synchronization is just one specific occasional coupling scheme. Other occasional coupling schemes can also in principle impart measure synchronization to a measure desynchronized state. We have checked that the transient uncoupling scheme certainly works. However, it may not be immediately obvious if a simple analytical prediction of the type given by Eq.~(\ref{eq:eff}) exists for other methods as well so that rather than implementing the schemes in an \emph{ad hoc} manner, one can find the optimal conditions beforehand in order to overcome the measure desynchronization.

Another potentially interesting direction of research could be to study the Kuramoto dynamics in the Hamiltonian systems~\cite{witthaut14} with a view to finding the relationship between the measure synchronization and the phase synchronization, and the effect of the occasional uncoupling on it. Last but not the least, in view of the recent extension~\cite{qiu2014} of the concept of the measure synchronization to address an analogous synchronization between two quantum many-body systems, it might be interesting to investigate the effect of the occasional coupling schemes in such systems.
%
\section*{Acknowledgements}
The authors thank Manu Mannattil for fruitful discussions, Shailendra K. Rathor for help with the numerics, and the two anonymous referees for the comments resulting in the improvement in the presentation of the paper. S.C. gratefully acknowledges financial support from the INSPIRE faculty fellowship $(\text{DST}/\text{INSPIRE}/04/2013/000365)$ awarded by the INSA, India and DST, India.
\bibliography{Ghosh_etal_manuscript.bib}
\end{document}